\documentclass[article,12pt,superscriptaddress,showkeys,showpacs] {revtex4}
\usepackage{amsfonts}
\usepackage{graphicx, amsmath}
\begin{document}

\title[Short Title]{Optically controlled phase gate and teleportation of a controlled-NOT gate for spin qubits in quantum dot-microcavity coupled system}
\author{Hong-Fu Wang\footnote{E-mail: hfwang@ybu.edu.cn}}
\affiliation{Department of Physics, College of Science, Yanbian
University, Yanji, Jilin 133002, People's Republic of China}
\affiliation{School of Physics and Optoelectronic Technology, Dalian University of Technology, Dalian 116024,
People's Republic of China}
\author{Ai-Dong Zhu}
\affiliation{Department of Physics, College of Science, Yanbian
University, Yanji, Jilin 133002, People's Republic of China}
\author{Shou Zhang}
\affiliation{Department of Physics, College of Science, Yanbian
University, Yanji, Jilin 133002, People's Republic of China}
\author{Kyu-Hwang Yeon}
\affiliation{Department of Physics \& BK21 Program for Device
Physics, College of Natural Science, Chungbuk National University,
Cheongju, Chungbuk 361-763, Republic of Korea}
\begin{abstract}
Assisted with linear optical manipulation, single photon, entangled photon pairs, photon measurement, and classical communication, a scheme for two-spin qubits phase gate and teleportation of a CNOT gate between two electron spins from acting on local qubits to acting on remote qubits using quantum dots in optical microcavities is proposed. The scheme is based on spin selective photon reflection from the cavity and is achieved in a deterministic way by the sequential detection of photons and the single-qubit rotations of a single electron spin in a self-assembled GaAs/InAs quantum dot. The feasibility of the scheme is assessed showing that high average fidelities of the gates are achievable in the weak-coupling regime when the side leakage and cavity loss are low. The scheme opens promising perspectives for long-distance quantum communication, distributed quantum computation, and constructing remote quantum information processing networks.
\pacs {03.67.-a, 42.50.Pq, 78.67.Hc}
\keywords{phase gate, controlled-NOT gate, quantum dot}
\end{abstract}

\maketitle \section{Introduction}\label{sec0}
Nowadays quantum computation, which could dramatically speed up the solution of certain mathematical problems, has developed into a truly interdisciplinary field owing to the contributions of physicists, computer scientists, and engineers. Moreover, various physical realizations of quantum computations are intensively studied, such as in cavity quantum electrodynamics (QED) system~\cite{QCWHHPRL9575, AGSPMJSPRL9983, XYSYGPRA0775}, linear optics
system~\cite{XSKGPRA0775}, ion trap system~\cite{JPPRL9574, DCDN02417}, nuclear magnetic resonance (NMR)
system~\cite{NIS97275}, Josephson charge qubits in superconducting circuit~\cite{AJMPRL0390},
and so on. The key ingredients of achieving quantum computation are quantum logic gates. A universal quantum computation network can be constructed from only two types of gates: one is single-qubit unitary gate and the other one is two-qubit conditional phase gate or controlled-NOT (CNOT) gate. Any unitary transformation, including multiqubit gates, can be decomposed into these elementary quantum gates in principle~\cite{ACRDNPTJHPRA9552}. On the other hand,
teleportation of quantum gates, which is viewed as a quantum remote control for the optimal implementation of nonlocal quantum operations by using local operations, classical communication, and prior shared entanglement (LOCCPSE) (that is, an unknown quantum gate operation acting on the local system is teleported and acts on an unknown state belonging to the remote system without physically sending the device), is a crucial way for constructing quantum information processing networking and teleportation-based building blocks of quantum communication and quantum computation. The general ideas for teleportation of quantum gates have been detailedly discussed in Refs.~\cite{MIPRL9779, AKPRA9858, JKPMPRA0062, DNSPRA0164}. It has been found that to implement a nonlocal CNOT gate with the help of LOCCPSE, it needs to consume only one ebit (maximally entangled pairs of qubits) entanglement and one cbit (bits of classical communication) in each direct~\cite{JKPMPRA0062}. A series of schemes have been proposed to investigate how to implement some special quantum gate operations~\cite{SMJPRA0265, YYLGCP0219, MMGJMO0350} and arbitrary unitary gate operations~\cite{ACSPRA0163, SJAMPRA0163, WGJPRL0289} by using entangled state as quantum channel.

Recent developments in semiconductor nanoelectronics technology have shown that semiconductor quantum dots (QDs), hailed for their potential scalability, are promising candidates as qubits for solid-state-based quantum-information processing and quantum computation because semiconductor QDs have properties similar to those of atoms such as discrete energy levels, coherent optical properties, and controllable coherent quantum evolution. The attractive advantage of storing the logical values in spin states is their relative isolation from the environment and the relatively long coherence time measured for single electron spin. Furthermore, the charge dynamics dependent on the spin state of an
electron could be optically induced via the Pauli exclusion principle and optical selection rules. Much effort has been dedicated to investigating fast initialization of the spin state of a single electron~\cite{CXDSLPRL0798, XYBQJDADCLPRL0799}, fast spin nondestructive measurement~\cite{DSSMAMTDPRL08101}, and fast optical control and coherent manipulation of a QD spin~~\cite{DTBYN08456, EKXBDADLPRL10104} and to demonstrating the implementations of optically controlled single-bit rotation gate and two-bit quantum phase gate for spin qubits in QDs~\cite{CLJPCM0719, DDPRA9857, ADGDDMAPRL9983, CFSJDMDPRL10104, TVLPRB1183, CJPRB1183, RYTMTKTYSPRL11107, LLEPRB1285}.

Recently, Bonato {\it et al.}~\cite{CFSJDMDPRL10104} demonstrated an interesting work showing that a single-electron-charged QD in the weak-coupling cavity QED regime exhibited a good interaction between a photon and an electron spin. Based on spinselective photon reflection from the cavity, the hybrid entanglement and CNOT gate between a photon and an electron spin could be efficiently realized with a QD coupled to a microcavity, and the complete two-photon Bell-state analyzer processes are also discussed. Based on giant circular birefringence, very recently, Hu and Rarity~\cite{CJPRB1183} proposed efficiently loss-resistant schemes for heralded state teleportation and entanglement swapping using a charged QD carrying a single spin coupled to an optical
microcavity. They discussed the fidelity and efficiency in the weak and strong coupling regimes, respectively, concluding that these schemes could be realized with current technology. In this paper, inspired by the above works, we propose an efficient scheme to implement a two-spin qubits phase gate and to teleporte a CNOT gate between two electron spins from acting on local qubits to acting on remote qubits, resorting to linear optical manipulation, single photon, entangled photon pairs, photon measurement, and classical communication. The scheme operates in the weak-coupling cavity QED regime and is achieved in a heralded way by the sequential detection of photons and the single-qubit rotations of a single electron spin in a self-assembled GaAs/InAs QD. The proposed scheme is simple and feasible as only single-spin rotation and single-photon detection are required, and it may open promising perspectives for long-distance quantum communication, distributed quantum computation, and constructing remote quantum information processing networks.

The paper is organized as follows. In Sec.~II, we first introduce the construction of a spin-cavity unit, then we illustrate how to implement an optically controlled phase gate and teleport a CNOT gate for spin qubits based on this spin-cavity system. In Sec.~III, we analyze and discuss the experimental challenge for the present scheme. A conclusion is given in Sec.~IV.

\begin{figure}
\includegraphics[width=4.0in]{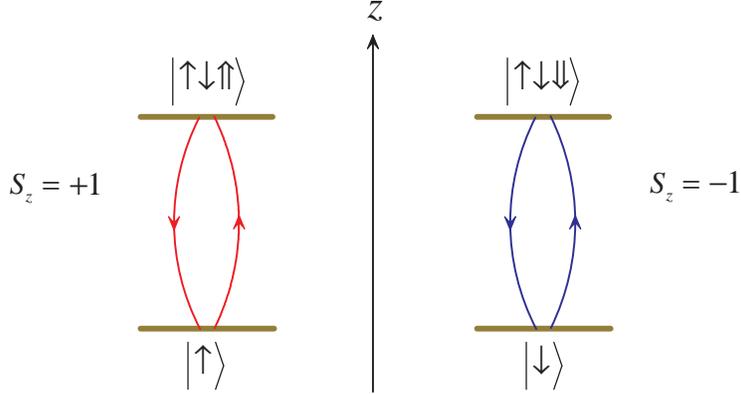}\caption{(Color online) Relevant energy level and optical selection rules for the optical
transition of $X^-$. Here the superscript arrow is to indicate their propagation
direction along the $z$ axis.}
\end{figure}

\section{Optically controlled phase gate and teleportation of a CNOT gate for spin qubits}\label{sec1}
We consider a singly charged GaAs/InAs QD, which has four relevant
electronic levels $|\uparrow\rangle$, $|\downarrow\rangle$, $|\uparrow\downarrow\Uparrow\rangle$, and $|\uparrow\downarrow\Downarrow\rangle$, as shown in Fig.~1, being embedded in a microcavity. The optical
excitation of the system will produce an exciton with negative charges and the charged exciton consists of
two electrons bound in one hole. According to the optical selection rules: the left circularly
polarized photon only couples the transition from the electron in the spin-up state $|\uparrow\rangle$ ($|\frac{1}{2}\rangle$) to the exciton $X^-$ in the trion state $|\uparrow\downarrow\Uparrow\rangle$; the right circularly polarized photon only couples the transition from the electron in the spin-down state $|\downarrow\rangle$ ($|-\frac{1}{2}\rangle$) to the trion state $|\uparrow\downarrow\Downarrow\rangle$, where $|\Uparrow\rangle=|\frac{3}{2},\frac{3}{2}\rangle$ and $|\Downarrow\rangle=|\frac{3}{2},-\frac{3}{2}\rangle$
represent heavy hole states with spin $3/2$ and $-3/2$ components. The trion state consists
of two electrons in a singlet state and a heavy hole, with the two trion levels being
\begin{eqnarray}\label{e1}
|\uparrow\downarrow\Uparrow\rangle&=&\frac{1}{\sqrt{2}}(|\uparrow\downarrow\rangle-|\downarrow\uparrow\rangle)|\Uparrow\rangle,\cr\cr
|\uparrow\downarrow\Downarrow\rangle&=&\frac{1}{\sqrt{2}}(|\uparrow\downarrow\rangle-|\downarrow\uparrow\rangle)|\Downarrow\rangle,
\end{eqnarray}
which indicates that the two electrons in a trion state have total spin zero, leading to that the electron-spin interactions with the heavy hole spin are avoided. In the limit of a weak incoming field, as illustrated in Ref.~\cite{ACJJPRA0775}, the electron-spin-cavity system behaves like a beam splitter. Based on the transmission and reflection rules of the cavity for an incident circular polarization photon with $s_z=\pm 1$ conditioned
on the QD-spin state, the dynamics of the interaction between photon and electron
in QD-microcavity coupled system is described as below~\cite{CFSJDMDPRL10104}:
\begin{eqnarray}\label{e2}
&&|R^\uparrow, \uparrow\rangle\rightarrow|L^\downarrow, \uparrow\rangle,~~~~~~~~~~~~|L^\uparrow, \uparrow\rangle\rightarrow -|L^\uparrow, \uparrow\rangle,\cr\cr&&|R^\downarrow, \uparrow\rangle\rightarrow -|R^\downarrow, \uparrow\rangle,~~~~~~~~~~|L^\downarrow, \uparrow\rangle\rightarrow |R^\uparrow, \uparrow\rangle,\cr\cr&&|R^\uparrow, \downarrow\rangle\rightarrow -|R^\uparrow, \downarrow\rangle,~~~~~~~~~~|L^\uparrow, \downarrow\rangle\rightarrow |R^\downarrow, \downarrow\rangle,\cr\cr&&|R^\downarrow, \downarrow\rangle\rightarrow |L^\uparrow, \downarrow\rangle,~~~~~~~~~~~~~|L^\downarrow, \downarrow\rangle\rightarrow -|L^\downarrow, \downarrow\rangle,
\end{eqnarray}
where $|L\rangle$ and $|R\rangle$ denote the states of the left- and right-circularly-polarized photons, respectively. The superscript arrow in the photon state indicates the propagation direction along the $z$ axis, and the arrows denote the direction of the electrons.

We now show how to implement a quantum phase gate of two spin qubits using the photon-spin interaction rules discussed above. The schematic is shown in Fig.~2. An input photon is in the polarization state $|L\rangle$, the electron spin 1 is in the state $a|\uparrow\rangle_1+b|\downarrow\rangle_1$, and the electron spin 2 is in the state $c|\uparrow\rangle_2+d|\downarrow\rangle_2$. The left-circularly-polarized photon first passes through a polarizing beam splitter in the circular basis ($c$-PBS), which transmits the input right-circularly-polarized photon $|R\rangle$ and reflects the left-circularly-polarized photon $|L\rangle$. Then the photon is injected into the first optical microcavity interacting with the first QD spin (Spin 1). The state evolution of the photon-spin system is given by
\begin{figure}
\includegraphics[width=4.0in]{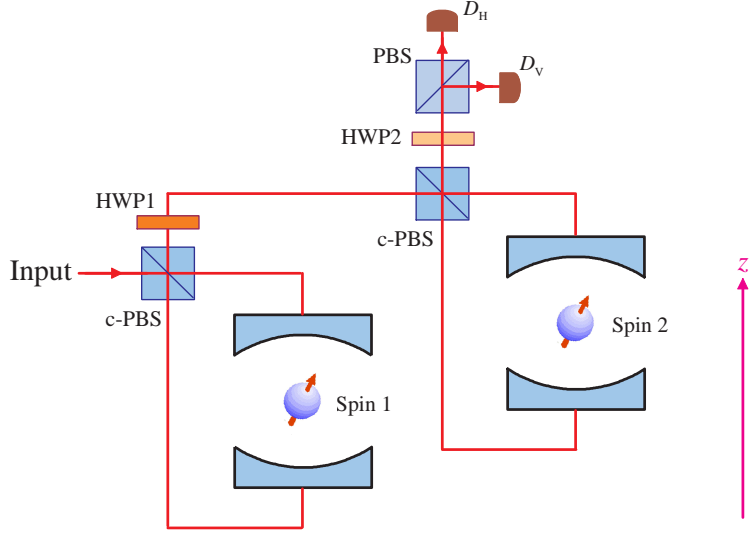}\caption{(Color online) Schematic of quantum phase gate for two quantum dot spins. Spin 1 and Spin 2 denote two QD spins coupled with two optical microcavities, respectively. The $z$ axis denotes the spin direction of the electron. $c$-PBS denotes the polarizing beam splitter in the circular basis, PBS denotes polarization beam splitter, which transmits the horizontal polarization $|H\rangle$ and reflects vertical polarization $|V\rangle$, HWP$i$ are half-wave plates, and $D_H$ and $D_V$ are conventional photon detectors.}
\end{figure}
\begin{eqnarray}\label{e3}
|\psi\rangle_0=-(a|\uparrow\rangle_1|L^\uparrow\rangle-b|\downarrow\rangle_1|R^\downarrow\rangle)(c|\uparrow\rangle_2+d|\downarrow\rangle_2).
\end{eqnarray}
After either transmitted or reflected by the optical cavity, the photon components are incident on the $c$-PBS again.
Next, the photon is rotated by a half-wave plate (HWP1), whose action is given by the transformation $|L\rangle\rightarrow (-1/\sqrt{2})(|L\rangle-|R\rangle)$ and $|R\rangle\rightarrow (1/\sqrt{2})(|L\rangle+|R\rangle)$, giving
\begin{eqnarray}\label{e4}
|\psi\rangle_1=-\frac{1}{\sqrt{2}}{\big[}a|\uparrow\rangle_1(|R\rangle-|L\rangle)
-b|\downarrow\rangle_1(|R\rangle+|L\rangle){\big]}(c|\uparrow\rangle_2+d|\downarrow\rangle_2).
\end{eqnarray}
Then the photon passes through another $c$-PBS and enters the second optical microcavity interacting with the second QD spin (Spin 2). We obtain
\begin{eqnarray}\label{e5}
|\psi\rangle_2=\frac{1}{\sqrt{2}}{\big[}a|\uparrow\rangle_1|(c\uparrow\rangle_2+d|\downarrow\rangle_2)(|R^\downarrow\rangle-|L^\uparrow\rangle)
-b|\downarrow\rangle_1(c|\uparrow\rangle_2-d|\downarrow\rangle_2)(|R^\downarrow\rangle+|L^\uparrow\rangle){\big]}.
\end{eqnarray}

After transmission from the second optical microcavity, the photon components are incident on the $c$-PBS again, and then it passes through another half-wave plate HWP2, which is used to complete the transformation between the linear polarization and the circular polarization, $|L\rangle\leftrightarrow|V\rangle$ and $|R\rangle\leftrightarrow|H\rangle$. The resulting evolution of the photon-electron state is written as
\begin{eqnarray}\label{e6}
|\psi\rangle_r&=&\frac{1}{\sqrt{2}}{\big\{}{\big[}a|\uparrow\rangle_1{\big(}c|\uparrow\rangle_2+d|\downarrow\rangle_2{\big)}-b|\downarrow\rangle_1{\big(}c|\uparrow\rangle_2
-d|\downarrow\rangle_2{\big)}{\big]}|H\rangle\cr\cr&&
-{\big[}a|\uparrow\rangle_1{\big(}c|\uparrow\rangle_2+d|\downarrow\rangle_2{\big)}+b|\downarrow\rangle_1{\big(}c|\uparrow\rangle_2-d|\downarrow\rangle_2{\big)}{\big]}|V\rangle{\big\}}.
\end{eqnarray}
After the photon polarization measurement and appropriate single-qubit gate rotation on electron spins, a two-qubit phase gate for two electron spins,
\begin{eqnarray}\label{e7}
|\uparrow\rangle_1|\uparrow\rangle_2\rightarrow|\uparrow\rangle_1|\uparrow\rangle_2,~~~~~~ |\uparrow\rangle_1|\downarrow\rangle_2\rightarrow|\uparrow\rangle_1|\downarrow\rangle_2,~~
\cr\cr|\downarrow\rangle_1|\uparrow\rangle_2\rightarrow|\downarrow\rangle_1|\uparrow\rangle_2,~~~~~~ |\downarrow\rangle_1|\downarrow\rangle_2\rightarrow-|\downarrow\rangle_1|\downarrow\rangle_2,
\end{eqnarray}
is accomplished (see Table I).

\begin{table}
\caption{The correspondence to the measurement results of photon polarization, the projected states of electron spins 1 and 2, and the corresponding single-qubit operations on electron spins in the case of implementing quantum phase gate for two-spin qubits.}\label{0}
\begin{tabular}{ccc}\hline\hline
{~~Measurement results~~}& {~~~~~~~~~~~~~~~~Projected states of electron spins~~~~~~~~~~~~~~~~}  &{~~Operations~~}\\\hline
$|H\rangle$&$ac|\uparrow\rangle_1|\uparrow\rangle_2+ad|\uparrow\rangle_1|\downarrow\rangle_2-bc|\downarrow\rangle_1|\uparrow\rangle_2+bd|\downarrow\rangle_1|\downarrow\rangle_2$&
$\sigma_z^1\otimes I^2$\\
$|V\rangle$&$ac|\uparrow\rangle_1|\uparrow\rangle_2+ad|\uparrow\rangle_1|\downarrow\rangle_2+bc|\downarrow\rangle_1|\uparrow\rangle_2-bd|\downarrow\rangle_1|\downarrow\rangle_2$&
$I^1\otimes I^2$\\\hline\hline
\end{tabular}
\end{table}

In the following we turn to the problem of teleporting a CNOT gate between two remote electron spins, as shown in Fig.~3. Photons 1 and 2 are prepared in the state $(1/\sqrt{2})(|R\rangle_1|R\rangle_2+|L\rangle_1|L\rangle_2)$,
and electron spins 1 and 2 are initialized to the states $\alpha|\uparrow\rangle_1+\beta|\downarrow\rangle_1$ and $\gamma|\uparrow\rangle_2+\delta|\downarrow\rangle_2$, respectively. Here photon 1 and electron spin 1 are in Alice's site and photon 2 and electron spin 2 are in Bob's site. Bob first performs a Hadamard gate transform ($|\uparrow\rangle\rightarrow(|\uparrow\rangle+|\downarrow\rangle)/\sqrt{2}$ and $|\downarrow\rangle\rightarrow(|\uparrow\rangle-|\downarrow\rangle)/\sqrt{2}$) on electron spin 2, the photon 2 is rotated by a half-wave plate HWP1, giving
\begin{eqnarray}\label{e8}
|\varphi\rangle_0&=&\frac{1}{2\sqrt{2}}(\alpha|\uparrow\rangle_1+\beta|\downarrow\rangle_1)
{\big[}\gamma(|\uparrow\rangle_2+|\downarrow\rangle_2)+\delta(|\uparrow\rangle_2-|\downarrow\rangle_2){\big]}
\cr\cr&&\otimes{\big[}|R\rangle_1(|R\rangle_2+|L\rangle_2)+|L\rangle_1(|R\rangle_2-|L\rangle_2){\big]}.
\end{eqnarray}
Then photons 1 and 2 pass through a $c$-PBS and enter the optical microcavities interacting with the QD spins 1 and 2, respectively. We obtain
\begin{eqnarray}\label{e9}
|\varphi\rangle_1&=&\frac{1}{2\sqrt{2}}{\big\{}\alpha|\uparrow\rangle_1{\big[}\gamma(|\uparrow\rangle_2
-|\downarrow\rangle_2)+\delta(|\uparrow\rangle_2+|\downarrow\rangle_2){\big]}|R^\downarrow\rangle_1(|R^\downarrow\rangle_2
+|L^\uparrow\rangle_2)\cr\cr&&+
\alpha|\uparrow\rangle_1{\big[}\gamma(|\uparrow\rangle_2+|\downarrow\rangle_2)
+\delta(|\uparrow\rangle_2-|\downarrow\rangle_2){\big]}|L^\uparrow\rangle_1(|R^\downarrow\rangle_2-|L^\uparrow\rangle_2)
\cr\cr&&-\beta|\downarrow\rangle_1{\big[}\gamma(|\uparrow\rangle_2-|\downarrow\rangle_2)+\delta(|\uparrow\rangle_2
+|\downarrow\rangle_2){\big]}|L^\uparrow\rangle_1(|R^\downarrow\rangle_2+|L^\uparrow\rangle_2)\cr\cr&&-
\beta|\downarrow\rangle_1{\big[}\gamma(|\uparrow\rangle_2+|\downarrow\rangle_2)+\delta(|\uparrow\rangle_2
-|\downarrow\rangle_2){\big]}|R^\downarrow\rangle_1(|R^\downarrow\rangle_2-|L^\uparrow\rangle_2){\big\}}.
\end{eqnarray}
After that, photons 1 and 2 get away from optical microcavities and pass through a $c$-PBS and  a half-wave plate HWP2, respectively. Finally, a Hadamard gate transform is performed on electron spin 2. The resulting evolution of the total spin-photon system is given by
\begin{figure}
\includegraphics[width=5.0in]{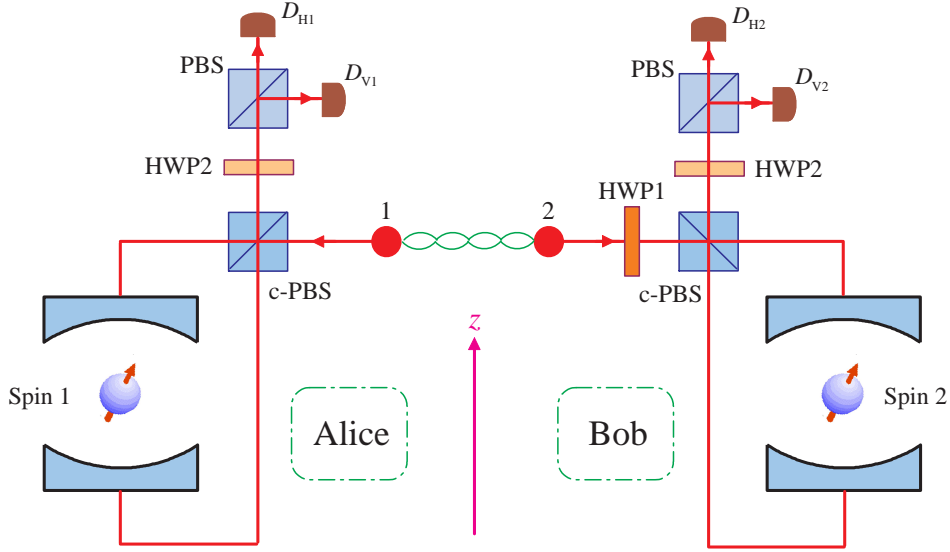}\caption{(Color online) Schematic of teleportation of a CNOT gate.
Here the function of half-wave plate HWP1 is to complete the transformation $|L\rangle\rightarrow (-1/\sqrt{2})(|L\rangle-|R\rangle)$ and $|R\rangle\rightarrow (1/\sqrt{2})(|L\rangle+|R\rangle)$, and HWP2 is to complete the transformation $|L\rangle\leftrightarrow|V\rangle$ and $|R\rangle\leftrightarrow|H\rangle$, respectively.}
\end{figure}
\begin{eqnarray}\label{e10}
|\varphi\rangle_r&=&\frac{1}{2}{\big[}(\alpha\gamma|\uparrow\rangle_1|\downarrow\rangle_2+\alpha\delta|\uparrow\rangle_1|\uparrow\rangle_2
-\beta\gamma|\downarrow\rangle_1|\uparrow\rangle_2-\beta\delta|\downarrow\rangle_1|\downarrow\rangle_2)|H\rangle_1|H\rangle_2\cr\cr&&+(\alpha\gamma|\uparrow\rangle_1|\downarrow\rangle_2+\alpha\delta|\uparrow\rangle_1|\uparrow\rangle_2
+\beta\gamma|\downarrow\rangle_1|\uparrow\rangle_2+\beta\delta|\downarrow\rangle_1|\downarrow\rangle_2)|H\rangle_1|V\rangle_2\cr\cr&&+(\alpha\gamma|\uparrow\rangle_1|\uparrow\rangle_2+\alpha\delta|\uparrow\rangle_1|\downarrow\rangle_2
-\beta\gamma|\downarrow\rangle_1|\downarrow\rangle_2-\beta\delta|\downarrow\rangle_1|\uparrow\rangle_2)|V\rangle_1|H\rangle_2\cr\cr&&-(\alpha\gamma|\uparrow\rangle_1|\uparrow\rangle_2+\alpha\delta|\uparrow\rangle_1|\downarrow\rangle_2
+\beta\gamma|\downarrow\rangle_1|\downarrow\rangle_2+\beta\delta|\downarrow\rangle_1|\uparrow\rangle_2)|V\rangle_1|V\rangle_2{\big]}.
\end{eqnarray}
After the photon polarization measurements and appropriate single-qubit gate rotation operations on electron spins, a nonlocal two-qubit CNOT gate for two electron spins,
\begin{eqnarray}\label{e11}
|\uparrow\rangle_1|\uparrow\rangle_2\rightarrow|\uparrow\rangle_1|\uparrow\rangle_2,~~~~~~ |\uparrow\rangle_1|\downarrow\rangle_2\rightarrow|\uparrow\rangle_1|\downarrow\rangle_2,
\cr\cr|\downarrow\rangle_1|\uparrow\rangle_2\rightarrow|\downarrow\rangle_1|\downarrow\rangle_2,~~~~~~ |\downarrow\rangle_1|\downarrow\rangle_2\rightarrow|\downarrow\rangle_1|\uparrow\rangle_2,
\end{eqnarray}
is accomplished (see Table II). Therefore, we achieve the teleportation of a CNOT gate between two remote electron spins successfully.

\begin{table}
\caption{The correspondence to the measurement results of photon polarizations, the projected states of electron spins 1 and 2, and the corresponding single-qubit operations on electron spins in the case of teleporting a CNOT gate between two remote electron spins.}\label{0}
\begin{tabular}{ccc}\hline\hline
{~~Measurement results~~}& {~~~~~~~~~~~~~~~~~Projected states of electron spins~~~~~~~~~~~~~~~~}  &{~~Operations~~}\\\hline
$|H\rangle_1|H\rangle_2$&$\alpha\gamma|\uparrow\rangle_1|\downarrow\rangle_2+\alpha\delta|\uparrow\rangle_1|\uparrow\rangle_2
-\beta\gamma|\downarrow\rangle_1|\uparrow\rangle_2-\beta\delta|\downarrow\rangle_1|\downarrow\rangle_2$&
$\sigma_z^1\otimes \sigma_x^2$\\
$|H\rangle_1|V\rangle_2$&$\alpha\gamma|\uparrow\rangle_1|\downarrow\rangle_2+\alpha\delta|\uparrow\rangle_1|\uparrow\rangle_2
+\beta\gamma|\downarrow\rangle_1|\uparrow\rangle_2+\beta\delta|\downarrow\rangle_1|\downarrow\rangle_2$&
$I^1\otimes \sigma_x^2$\\$|V\rangle_1|H\rangle_2$&$\alpha\gamma|\uparrow\rangle_1|\uparrow\rangle_2+\alpha\delta|\uparrow\rangle_1|\downarrow\rangle_2
-\beta\gamma|\downarrow\rangle_1|\downarrow\rangle_2-\beta\delta|\downarrow\rangle_1|\uparrow\rangle_2$&
$\sigma_z^1\otimes I^2$\\$|V\rangle_1|V\rangle_2$&$\alpha\gamma|\uparrow\rangle_1|\uparrow\rangle_2+\alpha\delta|\uparrow\rangle_1|\downarrow\rangle_2
+\beta\gamma|\downarrow\rangle_1|\downarrow\rangle_2+\beta\delta|\downarrow\rangle_1|\uparrow\rangle_2$&
$I^1\otimes I^2$\\\hline\hline
\end{tabular}
\end{table}

\section{Analysis and discussion}\label{sec2}
We now briefly analyze and discuss the feasibility and some practical issues we may face to for the implementation of the proposed scheme. Firstly, the gate fidelities in the present scheme mainly depend on the
coupling between the QD and microcavity system. Thus the key component in our scheme is the
spin-cavity unit, whose fidelity and efficiency have been discussed
detailedly in Ref.~\cite{CWJPRB0980}. As illustrated in Ref.~\cite{CWJPRB0980}, the frequency detuning and the normalized coupling strength played an important role to the performance of the spin-cavity
system. Generally, the reflection and transmission coefficients of the coupled (hot) and the uncoupled (cold) cavities are different when the side leakage and cavity loss are not negligible. In the weak excitation approximation, the reflection and transmission coefficients of a double-sided optical microcavity system is described by~\cite{CJPRB1183, CWJPRB0980, CAJWJPRB0878}
\begin{eqnarray}\label{e12}
r(\omega)&=&\frac{\left[i(\omega_{X^-}-\omega)+\frac{\gamma}{2}\right]\left[i(\omega_c-\omega)+\frac{\kappa_s}{2}\right]+g^2}{\left[i(\omega_{X^-}-\omega)+\frac{\gamma}{2}\right]\left[i(\omega_c-\omega)+\kappa+\frac{\kappa_s}{2}\right]+g^2},\cr\cr
t(\omega)&=&\frac{-\kappa\left[i(\omega_{X^-}-\omega)+\frac{\gamma}{2}\right]}{\left[i(\omega_{X^-}-\omega)+\frac{\gamma}{2}\right]\left[i(\omega_c-\omega)+\kappa+\frac{\kappa_s}{2}\right]+g^2},
\end{eqnarray}
where $g$ is the coupling strength, $\kappa$, $\kappa_s$, and $\gamma$ are the cavity field decay rate, leaky rate, and $X^-$ dipole decay rate, respectively. $\omega$, $\omega_c$, and $\omega_{X^-}$ are the frequencies of the input photon, cavity mode, and the spin-dependent optical transition, respectively. By setting $g=0$ and under the resonant interaction with $\omega_c=\omega_{X^-}=\omega_0$, the reflection and transmission coefficients for a cold
cavity with QD uncoupled to the cavity can be written as follows:
\begin{figure}
\includegraphics[width=4.5in]{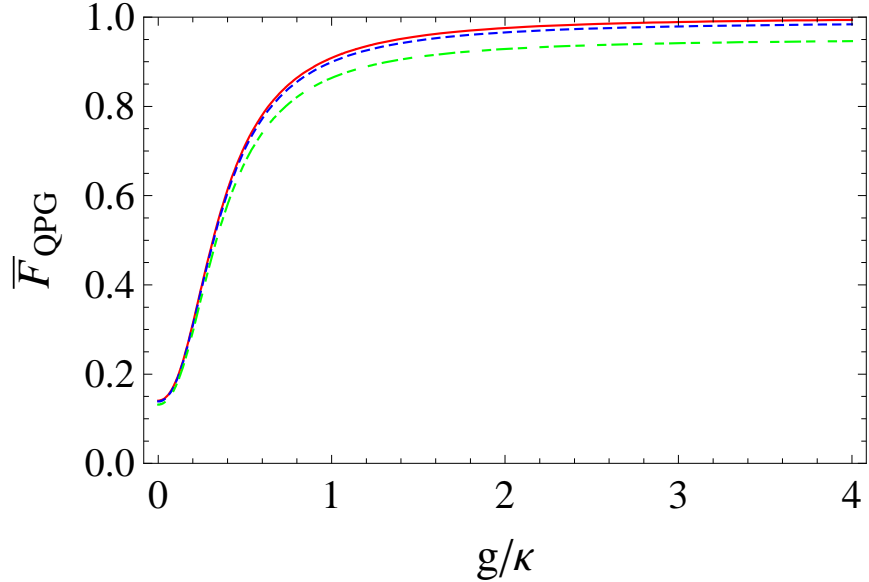}\caption{(Color online) The average fidelity of the two-spin quantum phase gate versus the normalized coupling strength $g/\kappa$. The solid line (red), dashed line (blue), and dotted-dashed line (green) corresponding to that the leaky rate is chosen as $\kappa_s=0$, $\kappa_s=0.01\kappa$, and $\kappa_s=0.05\kappa$, respectively. Here we have set that $\gamma=0.1\kappa$.}
\end{figure}
\begin{eqnarray}\label{e13}
r_0(\omega)&=&\frac{i(\omega_0-\omega)+\frac{\kappa_s}{2}}{i(\omega_0-\omega)+\kappa+\frac{\kappa_s}{2}},\cr\cr
t_0(\omega)&=&\frac{-\kappa}{i(\omega_0-\omega)+\kappa+\frac{\kappa_s}{2}}.
\end{eqnarray}
In the case of $\kappa\gg \kappa_s$ and $\omega=\omega_0$, the reflection coefficient of
the hot cavity $|r(\omega)|$ and transmission coefficient of the cold cavity $|t_0(\omega)|$ in the ideal case can approach unity. To qualify the performance of the phase gate and CNOT gate, we calculate the average fidelities of the two gates with respect to the coupling parameters, as shown in Fig.~4 and Fig.~5, which reveal that the cavity side leakage in the transmission process has a great impact on the gate fidelity with the increase of leaky rate $\kappa_s$. When $\kappa_s\ll\kappa$, we could make high gate fidelities even in the weakly coupling regime, which is easy to achieve experimentally. For the strong coupling, which is more challenging, has also been observed in various QD-cavity systems~\cite{N04432197, N04432200, PRL0595, APL0790}. Furthermore, the gate fidelities may be reduced by some small factors owing to the spin decoherence and trion dephasing, the same discussions as those in Ref.~\cite{CJPRB1183} are valid, which is not repeated here. Secondly, significant progress has recently been made
in the manipulation of single electron spins in QDs~\cite{DTBYN08456, S01292, S08320, NP095}. Schemes for fast initialization of the spin state of an electron and optically controlled single-qubit rotations for the spin of an electron in QDs have been demonstrated concretely in Refs.~\cite{CXDSLPRL0798, XYBQJDADCLPRL0799, DSSMAMTDPRL08101, DTBYN08456, EKXBDADLPRL10104, CLJPCM0719}. Therefore, the required preparation of electron-spin
superpositions and single-qubit gate rotation operations on the spin of an electron in our scheme can be effectively realized. Thirdly, indistinguishability and synchronization of photons are unnecessary as we exploit spin
coherence rather than photon coherence. Finally, the entangled state of photons used for implementing teleportation of a controlled-NOT gate can be produced with spontaneous parametric down-conversion (SPDC), a mature technology for the generation of entangled photon pairs. Moreover, the photon detectors used in our scheme are non-photon-number-resolving detectors that only can distinguish the vacuum and nonvacuum Fock number states, a sophisticated
single-photon detector distinguishing one or two photon states is
unnecessary. Although experiments for single-photon detectors have made tremendous progress, such
detectors still go beyond the current experimental technologies.
Our scheme thus greatly decreases the high-quality requirements of photon
detectors in practical realization.

\begin{figure}
\includegraphics[width=4.5in]{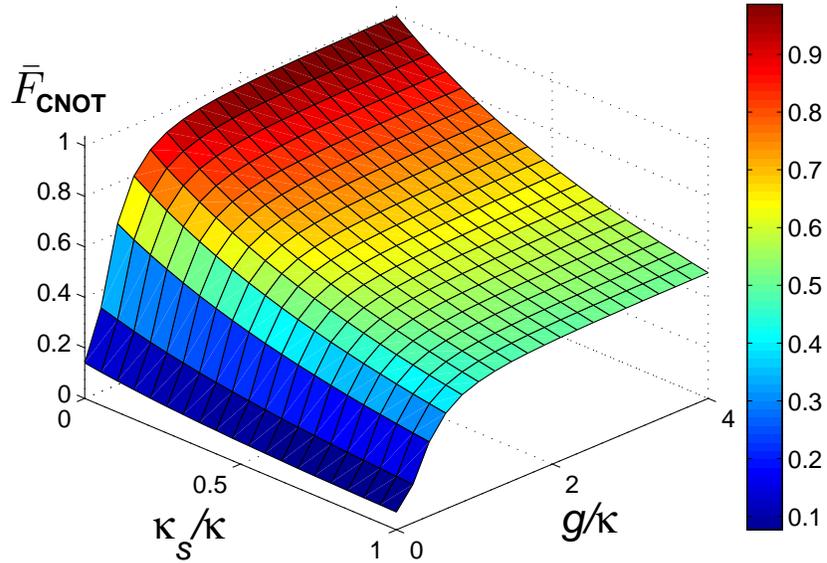}\caption{(Color online) The average fidelity of teleportation of the two-spin CNOT gate versus the normalized coupling strengths $\kappa_s/\kappa$ and $g/\kappa$. Here we have set that $\gamma=0.1\kappa$.}
\end{figure}

\section{Conclusions}\label{sec3}
In conclusion, we have proposed an efficient scheme for implementing a two-qubit optically controlled phase gate and teleporting a CNOT gate using QD spins in optical microcavities based on spin selective photon reflection from the cavity, resorting to linear optical manipulation, single photon, entangled photon pairs, photon measurement, and classical communication. In the ideal case, the scheme is deterministic, with high average fidelities when the side leakage and cavity loss are low. The phase gate and CNOT gate are heralded by the sequential detection of photons and the single-qubit rotations of a single electron spin. The scheme might be experimentally feasible with current technology and would open promising perspectives for long-distance quantum communication, distributed quantum computation, and constructing remote quantum information processing networks.

\begin{center}
{\small {\bf ACKNOWLEDGMENTS}}
\end{center}

This work is supported by the National Natural Science Foundation
of China under Grant Nos. 11264042, 61068001, and 11165015; China Postdoctoral
Science Foundation under Grant No. 2012M520612; the Program for Chun Miao Excellent Talents of Jilin Provincial Department of Education under Grant No. 201316; and the Talent Program of Yanbian University of China under Grant No. 950010001.

\end{document}